\documentclass[]{spie}  

\usepackage{caption}
\usepackage{subcaption}

\usepackage{amsmath,amsfonts,amssymb}
\usepackage{graphicx}
\usepackage[colorlinks=true, allcolors=blue]{hyperref}

\usepackage{mathtools}

\usepackage{txfonts}
\usepackage{adjustbox}

\title{Towards efficient machine-learning-based reduction of the cosmic-ray induced background in X-ray imaging detectors: increasing context awareness
}

\author[a]{Artem Poliszczuk}
\author[a]{Dan Wilkins}

\author[a,b,c]{Steven W. Allen}
\author[d]{Eric D. Miller}
\author[a]{Tanmoy Chattopadhyay}
\author[d]{Benjamin Schneider}
\author[a,b]{Julien Eric Darve}

\author[d]{Marshall Bautz}
\author[e]{Abe Falcone}
\author[d]{Richard Foster}
\author[d]{Catherine E. Grant}
\author[a]{Sven Herrmann}
\author[f]{Ralph Kraft}
\author[a,c]{R. Glenn Morris}
\author[f]{Paul Nulsen}
\author[a]{Peter Orel}
\author[f]{Gerrit Schellenberger}
\author[a,b]{Haley R. Stueber}

\affil[a]{Kavli Institute for Particle Astrophysics and Cosmology, Stanford University, 452 Lomita
Mall, Stanford, CA 94305, USA}
\affil[b]{Department of Physics, Stanford University, 382 Via Pueblo Mall, Stanford, CA 94305, USA}
\affil[c]{SLAC National Accelerator Laboratory, 2575 Sand Hill Road, Menlo Park, CA 94025, USA}
\affil[d]{MIT Kavli Institute for Astrophysics and Space Research, 77 Massachusetts Avenue, Cambridge, MA 02139, USA}
\affil[e]{Pennsylvania State University, Department of Astronomy and Astrophysics, University Park,
Pennsylvania, United States}
\affil[f]{Harvard–Smithsonian Center for Astrophysics, 60 Garden Street, Cambridge, MA 02138, USA}

\authorinfo{Send correspondence to Artem Poliszczuk\\E-mail: artempol@stanford.edu}

\begin{document} 
\maketitle

\begin{abstract}
Traditional cosmic ray filtering algorithms used in X-ray imaging detectors aboard space telescopes perform event reconstruction based on the properties of activated pixels above a certain energy threshold, within 3$\times$3 or 5$\times$5 pixel sliding windows. 
This approach can reject up to 98\% of the cosmic ray background. 
However, the remaining unrejected background constitutes a significant impediment to studies of low surface brightness objects, which are especially prevalent in the high-redshift universe. 
The main limitation of the traditional filtering algorithms is their ignorance of the long-range contextual information present in image frames. 
This becomes particularly problematic when analyzing signals created by secondary particles produced during interactions of cosmic rays with body of the detector. 
Such signals may look identical to the energy deposition left by X-ray photons, when one considers only the properties within the small sliding window. Additional information is present, however, in the spatial and energy correlations between signals in different parts of the same frame, which  
can be accessed by modern machine learning (ML) techniques . 
In this work, we continue the development of an ML-based pipeline for cosmic ray background mitigation. 
Our latest method consist of two stages: first, a frame classification neural network is used to create class activation maps (CAM), localizing all events within the frame; second, after event reconstruction, a random forest classifier, using features obtained from CAMs, is used to separate X-ray and cosmic ray features. 
The method delivers $>40$\% relative improvement over traditional filtering in background rejection in standard 0.3-10~keV energy range, at the expense of only a small ($<2$\%) level of lost X-ray signal. 
Our method also provides a convenient way to tune the cosmic ray rejection threshold to adapt to a user's specific scientific needs.

\end{abstract}


\section{INTRODUCTION}
\label{sec:introduction}

Astronomical imaging X-ray detectors aboard space-based instruments experience a significant cosmic-ray induced background~\cite{Campana22_inorbitXbkg}. 
While instruments located in low Earth orbits (below $\sim$2000 km) have some natural shielding to charged cosmic-ray particles from the geomagnetic field, instruments located at high Earth orbits and Lagrangian points typically suffer a substantial particle-induced background.
The charged particle background at the Lagrangian points has two main components: the solar wind\cite{solar_wind} and \textit{galactic cosmic rays} (GCR). 
The latter consists mainly of high-energy protons, with a smaller number of electrons and ions. 

Mitigating the impact of the cosmic ray background is crucial for the recovery of astronomical signals.
Current X-ray space observatories with pixelated imaging detectors implement shape-based grading schemes to identify signals produced by cosmic-rays. 
In most cases, the grading algorithm performs search through the science frames with a sliding 3$\times$3 or 5$\times$5 pixel window. 
Once the sliding window is centered on the most energetic pixel in the window, the pattern recognition of the pixel island shape is performed. 
Energy reconstruction is performed by summing the energies of all pixels in the window that are above a defined energy threshold. 
Such approaches provide an effective way to reduce the cosmic-ray background by up to ~98\%. 
However, the remaining $\sim 2$\% of unrejected cosmic ray background still poses a problem to studies of faint sources, such as outskirts of galaxies and galaxy clusters, and high redshift systems in general~\cite{ATHENA_white_paper_Nandra2013, Cluster_outskirts_Walker2019SSRv}. 

The unrejected cosmic ray background events are difficult to distinguish from the astrophysical X-ray photons due to the similarities of their pixel patterns. 
Interestingly, this type of event is frequently created by secondary particles produced during interaction of a primary cosmic ray with the body of the telescope and the spacecraft. 
Therefore, if the particle track of the primary particle is present in the frame, the spatial and energy correlations between the primary particle track and the secondary particle island in principle provide additional information that may be used to distinguish between valid X-ray and secondary particle islands. 
These spatial (and energetic) dependencies cannot be captured by traditional grading-based algorithms utilizing small sliding windows.

In this work we continue our investigation (Ref.~\citenum{WFI_bkg_vonKienlin2018SPIE, WFI_bkg_Grant2018SPIE, WFI_bkg_Bulbul2018SPIE, WFI_bkg_Grant2020SPIE, WFI_bkg_Eraerds2020SPIE, WFI_bkg_Miller2022JATIS, Wilkins2020,Wilkins2022, 
Poliszczuk23SPIE,
Wilkins24SPIE}) 
into how modern computational tools may be used to achieve more effective GCR background reduction with X-ray CCD and DEPFET sensors aboard future X-ray instruments, including the ATHENA/WFI~\cite{ATHENA_Barcons17, ATHENA_WFI_2016SPIE, ATHENA_WFI_2017SPIE} and AXIS~\cite{AXIS_Mushotzky2018SPIE}. 
In particular, we continue our study of machine learning (ML) based tools, as described in Ref.~\citenum{Wilkins2020, Wilkins2022, Poliszczuk23SPIE, Wilkins24SPIE}. 
In principle, machine learning methods should be able to incorporate the context of spatial, energy and grade correlations simultaneously, overcoming the limitations of traditional algorithms. 
The data used in this work are presented in Sec.~\ref{sec:data}.
Traditional filtering, machine learning models and evaluation methods are described in Sec.~\ref{sec:ml_methods}. Discussion of the ML model results and the comparison to traditional filtering methods can be found in Sec.~\ref{sec:results}. A summary is presented in Sec.~\ref{sec:summary}.

\section{Simulated training and validation data}
\label{sec:data}

The data for this work were created in a similar manner to our previous related studies.~\cite{Wilkins2020, Wilkins2022, Poliszczuk23SPIE, Wilkins24SPIE}. We use simulated cosmic ray data generated with the GEANT4~\cite{GEANT4_2003NIMPA, GEANT4_2006ITNS} software, employing a simplified mass model for the Wide Field Imager (WFI) instrument that will fly on the Athena X-ray Observatory~\cite{WFI_bkg_vonKienlin2018SPIE, WFI_bkg_Miller2022JATIS}. 
The WFI sensor has been approximated by four 450~$\mu$m thick silicon wafers, each of which is divided into 512$\times$512 pixels, of 130$\times$130~$\mu$m area each. 
The cosmic ray (CR) data were generated by randomly simulating 10$^9$ protons with an energy spectrum based on the CREME96~\cite{CREME96_1997ITNS} model for solar minimum. 
This set of CR primary particles was then simulated at the surface of 70~cm radius sphere surrounding the detector mass model. 
We will refer to the primary proton and all the secondary particles created during its interaction with the body of the detector as \textit{CR event}.
The X-ray photon events were simulated separately. We created 2$\times$10$^6$ X-ray events with the energies drawn from a log-uniform distribution in the $[0.2, 10.0]$~keV range.  
For the X-ray events, the depth of interaction was sampled from an exponential distribution with scale parameter equal to the mean interaction depth as a function of photon energy in silicon. 
The spatial position of a photon within a pixel was drawn from the uniform distribution. 

Position of X-ray interactions as well as all the interaction points of GCR events passing the silicon wafer were recorded on the 3-dimensional grid with 1$\mu$m resolution. 
To implement charge diffusion in our data, we applied linear analytical approximation from Ref.~\citenum{Iniewski07, Veale2014} to each of the interaction points on the grid. Here we assume no non-linearity of the electric field near the readout gates. 
For further description of the dataset creation and charge diffusion simulation see Ref.~\citenum{Poliszczuk23SPIE, Wilkins24SPIE}.
Once diffused, electron clouds from each interaction point of CR event (and a single interaction point for X-ray event) are sorted into the central and neighboring pixels. 
The deposited energy in each of the activated pixels is then corrected for the Fano energy resolution~\cite{Fano1947PhysRev} of silicon. 

To create a final dataset for the model training and validation we need to sort diffused events into frames. 
Following our approach from Ref.~\citenum{Wilkins2020, Wilkins2022, Poliszczuk23SPIE}, we use 64$\times$64 pixel sub-frames to train our model. 
This size of the frame seems to be an optimal trade-off between ability to identify long-range spatial correlations between different parts of the CR event and ability to train and test deep learning models in a short period of time. 
The 64$\times$64 sub-frame size allows one to capture~$\sim$86\% of CR particle showers. The distribution of particle shower size above 64 pixels flattens and no specific larger scale was found in our simulated dataset (for details see Ref.~\citenum{Poliszczuk23SPIE}). 

Dataset of frames was created by sampling CR and X-ray events and placing them on the 64$\times$64 pixel frame with spatial localization sampled at random. 
Once sampled events were placed in the frame, 
we added seven electron Gaussian readout noise to the frame. 
To get better results we transformed data representation from frames with electrons deposited in pixels into scaled logarithm of energy with zero mean and unit variance (see Ref.~\citenum{Poliszczuk23SPIE} for further discussion).
In our dataset we avoided pile-ups by imposing requirement of each pixel being activated by either X-ray or CR and not by both during sampling of the spatial localization of events. 
Number of events per frame was sampled from Poisson distribution with mean event rate of one X-ray photon and one CR event per 512$\times$512 pixel quadrant.
This rate corresponds to the 5~ms frame readout of ATHENA/WFI instrument pointing towards low surface brightness source. 

In our method we used two stages of ML algorithms (see Sec.~\ref{sec:ml_methods}). 
For the first stage we used 
64$\times$64 
pixel frame classification model which was performing four-class classification. The classes were empty frame, photon-only frame, CR-only frame and mixed frame containing both X-ray and CR. 
To obtain proper rations of events on the 64$\times$64 sub-frames we created 5000 frames of the size 512$\times$512 pixels and cut them into 64$\times$64 patches.
Assuming rate of CR and X-ray events as described above, the mixed frames would constitute only ~2\% of the dataset.
To avoid such significant class imbalance we created equal number of frames for each class. In non-empty frames we preserved the ratio of events from the segmented 512$\times$512 dataset. For CR-only and X-ray only frames, 99\% of frames contained one event per frame, 0.99\% contained two events per frame. Remaining 0.01\% contained more than two events per frame. 
Mixed frames subset contained 98\% of frames with one X-ray and one CR event per frame. One percent of the remaining 0.02 part of the mixed frames subset contained more than three events per frame.
For the second stage we used classification model that operated on reconstructed events with energy between 0.3 and 10 keV (see Sec.~\ref{sec:second_stage}). 
Here we performed separation between X-ray and CR events. In the energy range of interest, CR events make only 0.16\% of the dataset, leaving a significant class imbalance in the second stage dataset.

To create a training dataset we used 
10$^6$ simulated X-rays and 5,434,396 CR events. 
This resulted in 200,000 frames with 50,000 frames per class.
The validation sets were generated from 
10$^6$ simulated X-rays and 2,385,463 CR events. 
Here we created two different sets $-$ fist one of them consisted of 60,000 frames and was used for the performance evaluation of the first stage algorithm. We will refer to this set as \textit{small validation dataset}. It is a subset of 45,000 non-empty frames that was also used to create receiver operating characteristic curves (see Sec.~\ref{sec:eveluation_rf}) and find probability threshold value (see Sec.~\ref{sec:results}). 
Second validation set consisted of 840,000 frames with 280,000 frames of each of non-empty classes. We will refer to this set as \textit{large validation dataset}. This set was used for the validation of the second stage algorithm.

\section{TRADITIONAL AND MACHINE LEARNING APPROACHES TO BACKGROUND REDUCTION}
\label{sec:ml_methods}

\subsection{Traditional filtering}
\label{sec:traditional_filtering}
In this work as the \textit{traditional filtering algorithm} we applied the most effective version of currently used methods similar to the very faint mode filtering implemented in Suzaku X-ray telescope~\cite{SUZAKU2007PASJ}. 
This method was based on 3$\times$3 pixel sliding window scanning the frame.
Once the sliding window was centered on the pixel with the highest energy in the 3$\times$3 region, the algorithm checked energy of the central pixel. 
If event energy value was below \textit{event threshold} it was rejected as invalid. 
In our approach we set up the event threshold value to 260~eV. 
This way we were able to identify diffused X-ray events with the lowest energy in our data (0.3 keV) and, at the same time, be significantly above the level of the readout noise. 
If the energy of the central pixel was above the event threshold, algorithm checked the shape created by activated pixels in the 3$\times$3 window. 
Pixels were considered activated when deposited energy is above the \textit{split threshold} set up to 80~eV. The chosen threshold corresponded to the 3$\sigma$ value of the readout noise energy distribution. 
Accepted shapes formed by the pixels above split threshold were singles, non-diagonal doubles, L-shaped triples and quads. 
If there were any activated pixels beside the central one (i.e. it was not a single) algorithm checked if there were any activated pixels around the 3$\times$3 window in the 5$\times$5 pixel surrounding centered at the same position. If there were any additional pixels with energy above split threshold, the event was rejected. 
Finally energy of the event was reconstructed by summing energy of pixels in 3$\times$3 window which were above split threshold. 
If the event with central pixel was above event threshold and valid shape had summed energy below \textit{minimum ionizing particle} (MIP) threshold, 
event was considered to be valid. 
In our work the MIP threshold was set to 10.0~keV.

\subsection{Weakly supervised object localization}

In our pipeline we use the neural network frame classifier as the model for \textit{weakly supervised object localization} (WSOL).
The \textit{weakly supervised learning} (WSL) problem is defined as the learning problem where model is provided only with partial information about the learning task~\cite{WSOL_survey_zhang22}. In the case of weakly supervised object localization we often train our model providing information about \textit{what} is present in the image, but not \textit{where} it is localized. 
Refraining oneself from providing more complete information to the model during training can be viewed as a rather peculiar choice. 
In general, model would benefit from an information that could help it to learn the problem more effectively, i.e. in the supervised setup to train a model for object localization one should provide a ground truth information about the localization in the training data. 
Weakly supervised setup is usually implemented when the more precise information is not available or is expensive to produce, e.g. one may have a large dataset of labeled images, however creation of additional segmentation masks for these images would require a significant amount of human input.

Our motivation to use weak supervision setup instead of the full supervision was rooted in the problem of class imbalance in the segmentation problem. 
The task of \textit{semantic segmentation} can be viewed as the pixel-wise classification, where to each pixel, a specific class label is assigned. 
A loss function for such problem reflects the ability of the model to properly assign such class labels. 
In our previous work~\cite{Poliszczuk23SPIE} 
we were assigning pixel labels of readout noise only, X-ray and cosmic ray. 
High sparsity of our image data creates a significant class imbalance and leads to difficult problem to learn.
While having 4096 or 262,144 pixels in the image for 64$\times$64 pixel frames and 512$\times$512 pixel frames respectively usually only 1 pixel is occupied by the X-ray event and only a few more by the cosmic ray. 
When calculating the average pixel misclassification after passing the frame through the neural network, wrong label assignment to a single X-ray pixel does not introduce a significant change to the general loss. Therefore information from the majority class, i.e. empty pixels filled with readout noise dominate the feedback model gets during training.
This property impose a major problem for the model if the problem we want to learn is the proper identification of X-ray signal. 
It also limits our ability to increase model's field of view, e.g. feed it with data from the full quadrant of the ATHENA/WFI detector.

In our previous work~\cite{Poliszczuk23SPIE}
we showed that we can effectively train a semantic segmentation model by applying a focal loss function~\cite{FocalLoss}, which focused on pixels that were difficult to classify. It allowed model to reduce the impact of empty, readout noise only pixels during training. While the model was able to converge and we were able to suppress the class imbalance problem we could not surpass the performance of traditional grading algorithms for background rejection. Our model was very effective in the recovery of the X-ray signal, but at the same time the detected signal contained a significant amount of unrejected cosmic-ray events. 
This behavior was reflected in the discrepancy between performance evaluation
made in pixel space and reconstructed event space.
The pixel space evaluation was done with standard segmentation metrics which allow one to analyze the performance of segmentation model by looking at how well pixels of specific class were classified. 
However, these metrics not always reflected how well the model output can be used for cosmic-ray background mitigation after event reconstruction (for details on event reconstruction procedure see Sec.~\ref{sec:traditional_filtering}).
Because of the different properties of the model output before and after event reconstruction, we treat these stages, pixel-level and event-level, as separate. 

Final classification of the event is done at the event-level stage, therefore the main role of the pixel-level model is class-related feature extraction from the frame. 
For this purpose we use a neural network model trained on a frame classification task. 
Such model similarly to frame classifier in~\citenum{Poliszczuk23SPIE} classifies frame as empty, X-ray only, CR-only or both (MIX).
Then we recover position and class of the event in form of class activation maps (CAM).
Finally we get reconstructed events and CAM-related features and pass them to the second stage of the filtering pipeline (see Sec.~\ref{sec:second_stage}).

The \textit{class activation map}~\cite{Zhou2016_CAM_main_paper} (CAM) can be viewed as the importance of the specific spatial location for the prediction of selected class of the image. 
When information is passing through the typical convolutional neural network (CNN) classifier each following layer is getting more general information connected to the final prediction task compared to the previous layer, loosing at the same time spatial information due to the pooling operations~\cite{max_pooling}. 
To get information related to the specific class we are using final convolutional layer of the network. 
Let $a_i(x, y)$ be the activation of the feature map $i$ in the last convolutional layer at spatial position $(x, y)$. 
Next we calculate average value of each of the feature maps called \textit{global average pooling} (GAP) which can be written as
$GAP_{i} = \sum_{x,y}a_i(x,y)$ for the feature map $i$.
The global average pooling was first introduced in~\citenum{Lin2013NetworkInNetwork_GAP} as a regularization technique which can be used instead of the fully connected layer at the end of the model. 
However, it was also discovered that it can be applied to object localization task in the weak supervision setup~\cite{Zhou2016_CAM_main_paper}.
When applying the GAP for the classifier model, the input of the softmax layer~\cite{softmax} for the class $k$ is
$S(y=k) = \sum_{i}w^{k}_{i}GAP_{i}$, 
where $w^{k}_{i}$ is the learned weight for the feature map $i$ for class $k$. Since class of the image is connected to spatial localization of the class object on the image, the weight $w^{k}_{i}$ can be viewed as the importance of the feature map $i$ for the class $k$ in a specific spatial location.
By combining above equations we get:

\begin{equation}
    S(y=k) = \sum_{i}w^{k}_{i}\sum_{x,y}a_i(x,y) = \sum_{x, y}CAM_{y=k}(x, y), 
\end{equation}

\noindent
where class activation map for class $k$ is defined as 
$CAM_{y=k}(x, y) = \sum_{i} w^{k}_{i} a_i(x,y)$. In other words, to obtain CAM for class $k$ we perform a class-wise weighted sum of activations from all feature maps of the last convolution layer for each of spatial locations. 

There is one additional property of the model one must keep in mind when using neural network for weakly supervised object localization: the final resolution of the model.
If the model is used solely for the classification purposes
pooling operations can be performed until there is no spatial information available. However, if the classification model is used for CAM-based object localization, there is a limit to the model depth: one cannot develop model with too many pooling layers.

As a frame classfication model we used
\textit{densely connected convolutional network} (DenseNet, \citenum{DenseNet_huang17}) which was found to be the best performing model in our previous work~\cite{Poliszczuk23SPIE}. 
The DenseNet is a modification of the more popular residual convolutional network~\cite{ResNet_He16}.
In the DenseNet, network is divided into separate convolutional blocks where each of the layers is connected to all previous layers in the block. 
This large number of skipped connections allows one to achieve better information flow through the network and reduce the number of parameters in the model.
Due to the necessity of keeping resolution of the final layer relatively large in order to perform CAM-based object localization we slightly modified previously used model.
We used one initial convolutional block and two dense blocks with five convolutional layers per dense block (see Ref.~\citenum{DenseNet_huang17} for details). 
The growth rate of the network was set to 12 with initial number of layers equal to 64. 
Each convolutional layer was built with the ReLU activation function~\cite{relu1, relu2}, batch normalization~\cite{BatchNorm_Ioffe15} and 0.2 dropout rate~\cite{DropoutPaper_srivastava14}.
The final resolution of the network was set to 16$\times$16 pixels. 
Class activation maps obtained from the DenseNet were then bilinearly upsampled to the original, 64$\times$64 pixel resolution.

\subsection{Second Stage: Event Reconstruction and Random Forest}
\label{sec:second_stage}

The second stage of the cosmic-ray filtering pipeline has two distinct parts.
First, event is reconstructed in a similar manner as it is done in the traditional algorithms~\footnote{Our team is currently developing more effective way to reconstruct energy and position of the incoming X-ray photon, which will be able to further improve the AI filtering model performance. For more details see~\citenum{Wilkins24SPIE}.}. 
Event reconstruction scheme described in Sec.~\ref{sec:traditional_filtering} is similar to the one used for the AI model. 
We are using the same energy thresholds as in the traditional filtering (split, event and MIP thresholds). We do not use any grade values to filter events - we analyze all the events with reconstructed energy being in the $[0.3, 10]$~keV range. 
When we scan the frame with 3$\times$3 pixel window and center it on the most energetic pixel in the window, we perform additional operations on class activation maps: if the energy of the event falls into the range of our interest, we also read minimum, maximum and median values in the corresponding 5$\times$5 window of X-ray and cosmic ray class activation maps. 
The 5$\times$5 pixel size of the CAM window was chosen because of the input-to-output resolution ratio of the neural network was set to four. 
In our early experiments we also tested incorporation of MIX class CAM, but this additional information did not improve the model performance. 
Measurements obtained from class activation maps were then combined into features used for the second stage classification. 
Details of their construction can be found in Sec.~\ref{sec:results}
Therefore the second stage classifier operates in the \textit{event-level space} with X-ray and cosmic ray CAM features instead of the \textit{frame-level space} which was used for the neural network frame classifier and \textit{pixel-level space} which was used for the semantic segmentation models in our previous work~\cite{Poliszczuk23SPIE}. 

In the event-level space we have a subset of events with valid energies. The label assigned to the frame (as being empty, X-ray only, cosmic ray only or mixed) can be a useful information for the second-stage classifier, but cannot be used as a label for the new model. Therefore we redefined the learning problem for the second stage classification as a binary classification between valid (X-ray) and invalid (cosmic ray) events. 
In the case of second stage we no longer have the equal size classes in the training and validation data. In fact, cosmic ray events with valid energies make only 1.3 percent of the dataset.
As described in Sec.~\ref{sec:performance_evaluation}, 
in the case of highly imbalanced data, it is common to assign 'positive' label to the smaller class. Therefore the second stage learning problem can be viewed as the recovery of cosmic ray events from the X-ray event dominated data.

We used the \textit{Random Forest} classification algorithm~\cite{breiman2001RandomForest} as a model for the second stage, event-level filtering. 
Random forest consists of ensemble of decision trees. 
Each decision tree is developed by minimizing impurity function which serves as the loss during training. 
The impurity function value is low when we are able to find a cut in the feature space that can improve separation between classes. 
In each node of the tree we perform a cut in a single feature. 
In our work we used Gini impurity function. 
The Gini impurity for the set $S$ with $K$ classes takes form of 
$G(S) = \sum_{k=1}^{K} p_k(1-p_k)$, 
where $p_k = |S_k| / |S|$ is the probability of a point from the set $S$ being a member of class $k$ and $S_k$ is the subset of $S$ containing $k$ class members. In the binary classification case it takes form of $2p(1-p)$, where $p$ is the probability of the positive class. 
When building a tree we calculate impurity in each leaf. If the leaf has zero loss, i.e. it contains members of only one class, we stop. If the leaf has large loss we split the leaf in two. 
In the default decision tree setup we search through all available features in each node, to find the most optimal split. 
This type of model, especially in the fully developed version, i.e. decision tree classifier containing only pure leaves, has a very high variance and easily overfits the data. 
The random forest algorithm solves this problem by incorporating two important components into the training. First one is \textit{bagging} also known as bootstrap aggregating. 
We create $m$ datasets out of initial dataset $D$. These are created by drawing elements of $D$ with replacement. Each of datasets $D_m$ has the same size as the original dataset $D$. 
Next we train a separate, fully developed decision tree classifier $t_m$ on each of the datasets $D_m$ (overfitting each separate $D_m$). Our final classifier $t$ is created by averaging predictions from each of $t_m$ trees.
Creation of multiple datasets out of $D$ allows to average out noisy properties of the whole dataset, because these properties will be different in different sets $D_m$. 
In this setup predicted class probabilities of an input sample $x$ are computed as the average predicted probabilities of the trees $t_m$ in the ensemble. The class probability of a single decision tree is the fraction of samples of the same class in a leaf.

An additional difficulty occurs when bagged tree ensemble is learning on a highly imbalanced data~\cite{chen04balanced_rf}. A bootstrap sample often may include very few or no elements of the minority class. This behaviour would result in a poor performance on the smaller class, and as a consequence, would lead to inability of the model to properly identify valid cosmic-ray events. 
To overcome this problem a class-wise bootstrap has to be introduced. In this work we use a solution described in~\citenum{chen04balanced_rf}: first we draw a bootstrap sample from the smaller class. Then we randomly draw with replacement the same number of elements from the bigger class. 

Second property that differs random forest from bagged trees ensemble is the reduced number of features tree can use in the split. Random forest allows to try only $k$ out of $d$ existing features. This feature subsample is randomly drawn from all-features set every time a decision tree makes a split.
Making this modification to the tree building process allows one to decorrelate decision trees in the ensemble and make it robust to overfitting.
The number of selected features $k$ is a model hyperparameter. However, it is a known practice to use $k = \lceil \sqrt{d} \rceil $, 
which we also apply. Second important parameter is the number of trees in the ensemble. In general larger number of trees gives one better results, until a certain number, after which adding new trees does not further improve the model performance.

\begin{table}[t]
\caption{Performance evaluation of the DenseNet frame classification model used in the first stage of our pipeline. All metrics values were multiplied by 100.}

\centering
    \begin{tabular}{c|c|c|c|c}
    \hline
    \textbf{Metric} & \textbf{EMPTY} & \textbf{X-RAY} & \textbf{GCR} 
    & \textbf{MIX} \\ 
    \hline
    F1        & 99.83 & 98.16 & 98.12 & 98.89 \\
    Precision & 99.67 & 97.21 & 99.48 & 98.57 \\
    Recall    & 100.0 & 99.13 & 96.81 & 99.21 \\
\hline
    \end{tabular}
    \label{tab:metrics_nn}
\end{table}

\begin{figure}[t]
    \centering
    \includegraphics[width=0.9\textwidth]{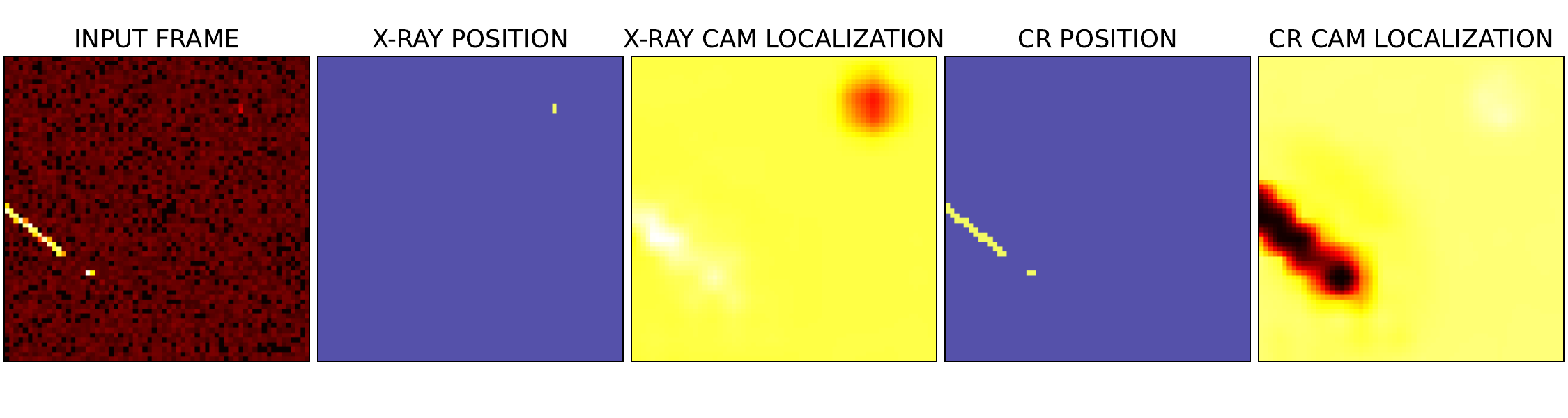}
    \caption{Object localization performed by the frame classification network via class activation maps (CAM). 
    The left panel shows the simulated frame which was passed as an input to the classification model. Second and fourth panels (presented with blue and yellow colors) are showing true positions of X-ray and cosmic ray signals respectively. 
    The third and fifth panels are showing localization capabilities of class activation maps.}
    \label{fig:cam_vis}
\end{figure}

\subsection{Performance evaluation}
\label{sec:performance_evaluation}

Evaluation of the model performance was done via analysis of various classification metrics. 
All of the used metrics can be constructed from the \textit{confusion matrix}, which shows the number of objects in the dataset having properly or improperly assigned a specific class label. The binary classification confusion matrix 
is usually separated into \textit{Positive} and \textit{Negative} classes, with \textit{True Positive} (TP) and \textit{True Negative} (TN) being properly classified positive and negative class members respectively. Sizes of these properly classified subsamples are located on the diagonal of the confusion matrix. Off-diagonal elements of the confusion matrix are occupied by the misclassification subsamples: these are \textit{False Positives} (FP), i.e. negative-class examples classified as positives and \textit{False Negatives} (FN) being the misclassified ground truth positive class members. 

When selecting metrics for the performance evaluation one needs to keep in mind the difference in classification problems we face at first and second stages of the algorithm. 
The first stage frame classifier deals with multi-class classification problem and is trained evaluated on the equal-size classes, i.e. we have the same number of empty, X-ray only, cosmic-ray only and mixed frames. 
To evaluate this stage we use four metrics: accuracy, precision, recall and F1 score.
Accuracy the metric showing ratio of properly classified objects to all objects present in the dataset. Equation~\ref{eq:accuracy} shows accuracy formula for the binary classification. Extension to multi-class version has the same structure:

\begin{equation}\label{eq:accuracy}
    \text{Accuracy} = \dfrac{\text{TP}+\text{TN}}{\text{TP} + \text{FP} + \text{TN} + \text{FN}}. 
\end{equation}

\noindent
While accuracy is widely used to evaluate model performance it may mislead one when used on the highly imbalanced dataset. In such case, if model misclassify minority class objects, it still would achieve a high accuracy.~\cite{imbalanced_metrics}
In the case of the dataset used to train and evaluate second stage algorithm the ratio of cosmic-ray events to X-ray events is around 1.5 percent. 
Due to such a highly imbalanced data we do not use accuracy in the second stage performance evaluation. 
In the following text, when discussing the second stage algorithm performance, we will refer to the smaller cosmic-ray event class as a positive class and to the large X-ray event class - as a negative class.

\noindent
\textit{Precision} is the ratio of properly classified positive class to the number of all predicted positive class members. In other words, precision can be viewed as a purity of a positive class:

\begin{equation}\label{eq:precision}
    \text{Precision} = \dfrac{\text{TP}}{\text{TP} + \text{FP}}. 
\end{equation}

\noindent
For the multi-class problem, purity of a specific class is defined by putting the selected class properties under the positive class label and all other classes - under the negative class. 
\textit{Recall} (or True Positive Rate, TPR) is identified as the completeness of the positive class, i.e. the ratio of properly classified positive class members to all existing positive class members (properly and improperly classified). For the multi-class problem, this metric is modified in the same way as precision. The binary classification recall can be written as:

\begin{figure}[t]
    \centering
    \includegraphics[width=0.49\textwidth]{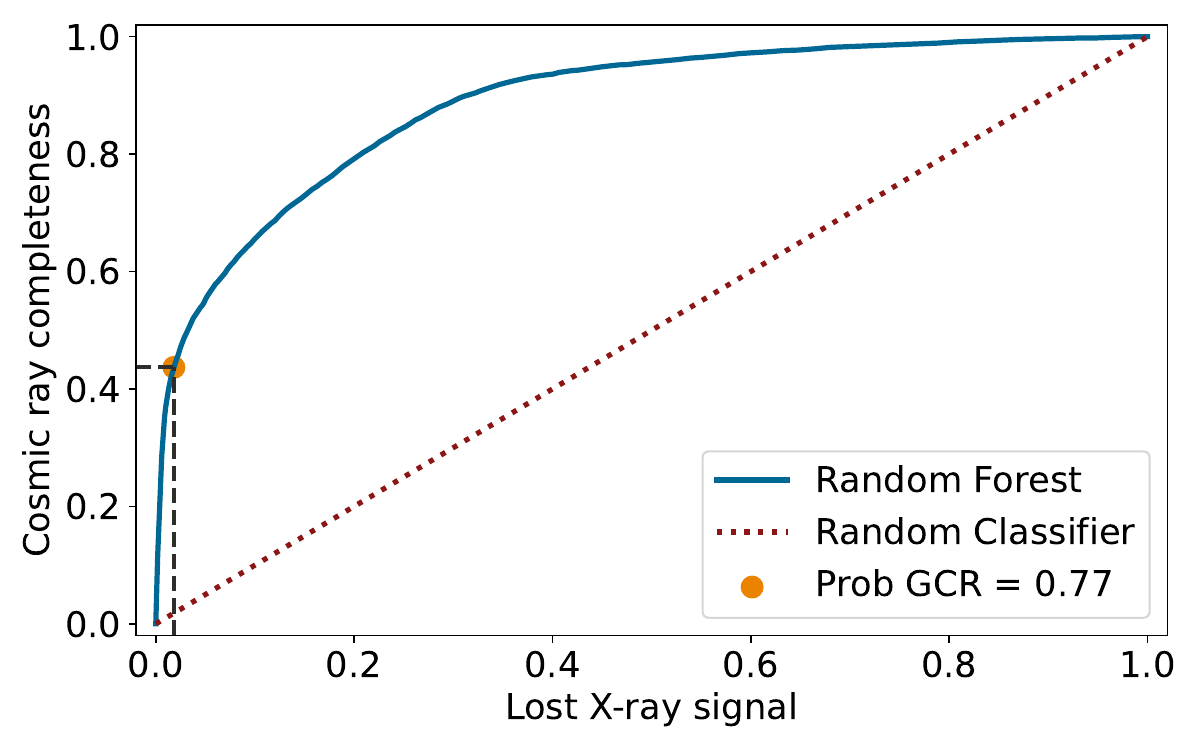}
    \caption{Receiver operating characteristic (ROC) curve for the random forest classifier (blue line). It shows pairs of cosmic ray class completeness and the fraction of lost X-ray signal for different probability thresholds. Threshold value of 0.77 (orange point) was chosen as an optimal one. It corresponds to 43.7\% of cosmic ray event completeness 1.82\% of X-ray lost signal. Red line shows ROC curve if cosmic ray events were selected at random.}
    \label{fig:roc_curve}
\end{figure}

\begin{equation}\label{eq:recall}
    \text{Recall} = \dfrac{\text{TP}}{\text{TP} + \text{FN}}.  
\end{equation}

\noindent
The \textit{F1-score}, shown in Eq.~\ref{eq:f1}, is a harmonic mean of precision and recall and is a reliable indicator of general model performance also in the case of class-imbalance problem. For the multi-class problem the class-specific F1-score is calculated by taking precision and recall for the chosen class.

\begin{equation}\label{eq:f1}
    \text{F1} = 2 \dfrac{\text{Precision} \cdot \text{Recall}}{\text{Precision} + \text{Recall}}.
\end{equation}

\noindent
While dealing with high class-imbalance, 
metrics such as F1-score, precision and recall gives us information about the general model classification capabilities as well as how well it can perform on the smaller class, we also need more task-specific metrics.
One has to remember that the final goal we want to achieve, is the reduction of cosmic-ray component in the detected X-ray signal, while keeping the ratio of lost astronomical X-rays to all X-rays sent by the source on the low level. 
Therefore, while dealing with the problem of much smaller cosmic-ray class in the second stage of the algorithm, we must carefully control the performance on the larger class as well.
Achieving both of these goals requires finding the optimal trade-off between rejecting all possible cosmic-ray candidates and keeping all X-ray events. 
One can do it by changing the probability threshold for identification of cosmic-ray event, provided by the second stage algorithm. 
Useful tool for finding such a threshold is the \textit{receiver operating characteristic} (ROC) curve, which is constructed by plotting points on the recall (true positive rate or completeness of the cosmic-ray sample) vs false positive rate for different probability thresholds. 
The \textit{false positive rate} (FPR) or \textit{lost X-ray signal} is defined as the ratio of misclassified X-ray events to the number of all events predicted to be X-rays:

\begin{equation}\label{eq:fpr}
    \text{Lost X-RAY signal} = \text{FPR} =  \dfrac{\text{FP}}{\text{TN} + \text{FP}} = \dfrac{\text{Misclassified X-RAYs}}{\text{All X-RAYs in the data}}.
\end{equation}

The ROC plot gives us the continuum of probability threshold values, allowing to find an optimal trade-off between aggressive filtering of all suspicious events leading to the very pure X-ray sample of very few events and ineffective filtering which gives us very high completeness of the detected X-ray signal with significant background component.
To have the better understanding of the cosmic-ray filtering algorithm performance we also use the \textit{unrejected cosmic-ray background} or \textit{false omission rate} (FOR) in terms of our negative and positive class definition. The FOR is defined as the ratio of cosmic-ray events present in our predicted X-ray signal to the size of the predicted X-ray signal:

\begin{equation}\label{eq:for}
    \text{Unrejected CR background} = \text{FOR} =  \dfrac{\text{FN}}{\text{TN} + \text{FN}} =   \dfrac{\text{Misclassified CRs}}{\text{Predicted X-RAYs}}.
\end{equation}

\noindent
Finally we also use the \textit{relative improvement in cosmic-ray background rejection} or \textit{FOR relative improvement}. It is defined as the difference in the unrejected background size between our model and a traditional filtering method:

\begin{equation}\label{eq:rel_imp}
    \text{FOR relative improvement} = 1 - \dfrac{\text{FOR(AI Filtering)}}{\text{FOR(Traditional Filtering)}}.
\end{equation}

\section{Results} \label{sec:results}

\subsection{Frame neural network classifier}

We trained our model for 100 epochs using Adam optimizer~\cite{Adam} with exponential learning rate decay with learning rate on $n$ epoch being $l_{n} = l_{0}d^n$. Initial learning rate $l_0$ was set to 0.007 and the decay rate $d$ was set to 0.95.
Our DenseNet model achieved an accuracy of 98.76. Additional metric values for specific classes are presented in Tab.~\ref{tab:metrics_nn}.
In order to preserve spatial information in the last convolutional layer for the CAM-based object localization, we had to reduce the depth of the network compared to the model presented in our previous work~\cite{Poliszczuk23SPIE}. 
This change could led to the decrease in the model performance.
However, obtained results do not differ in a significant way from values obtained in our previous work with deeper DenseNet~\cite{Poliszczuk23SPIE}. 
While implemented reduction of the depth of classification model did not affect model performance, more thorough study on the relation between the output resolution and classification performance need to be done, especially when it comes to the classification model operating on the full 512$\times$512 pixel frame.

Figure~\ref{fig:cam_vis} shows an example of CAM-based object localization. One can see a cosmic ray track as well as secondary island located at axis of cosmic ray track being properly identified and localized. The proper localization and class assignment can be also seen in the case of X-ray event. 
It is worth noting that activation map of a particular class shows larger values in the regions of avoidance, i.e. X-ray CAM is more suitable for cosmic ray localization and cosmic ray CAM - for X-ray localization respectively. 
We do not show MIX class CAM since introduction of the MIX CAM information into the random forest feature space did not improve classification performance.

\begin{table}[t]
\caption{Performance comparison of second stage ML algorithm (CAM and Random Forest) and traditional filtering algorithm in the 0.3-10~keV energy range. All metrics values were multiplied by 100.}

\centering
    \begin{tabular}{c|c|c}
    \hline
    \textbf{METRIC} & \textbf{AI FILTERING} & \textbf{TRADITIONAL FILTERING} \\ 
    \hline
    F1                         & 31.98 &  2.91 \\
    Precision                  & 25.23 &  3.31 \\
    Recall                     & 43.67 &  2.60 \\
    Lost X-RAY signal (FPR)     & 1.82  &  1.07 \\
    CR in detected signal (FOR)& 0.80  &  1.37 \\
    FOR relative improvement   & 41.37 & $-$\\
\hline
    \end{tabular}
\label{tab:classification_metrics_2nd_stage}
\end{table}

\subsection{Event classifier} \label{sec:eveluation_rf}

At the second stage of the cosmic-ray mitigation algorithm we perform binary classification between cosmic-ray and X-ray events. 
Here we operate at the level of reconstructed events instead of frames of certain class as it was in the case of the first stage algorithm.
After we performed event reconstruction on training dataset and large validation dataset we limited our data to events with energies in the $0.3-10.0$ keV range.
This way we have got 49,222 X-ray events and 660 GCR events for the training set and 458,875 X-ray and 6,464 GCR events for the validation set. 
It gives a significant class imbalance of GCR event class being only ~1.3\% of the dataset.
In our analysis we will often refer to the GCR event class as Positive and X-ray event class - as Negative.

\begin{figure}[b!]
     \centering
     \begin{subfigure}[b]{0.49\textwidth}
     \centering
     \includegraphics[width=\textwidth]{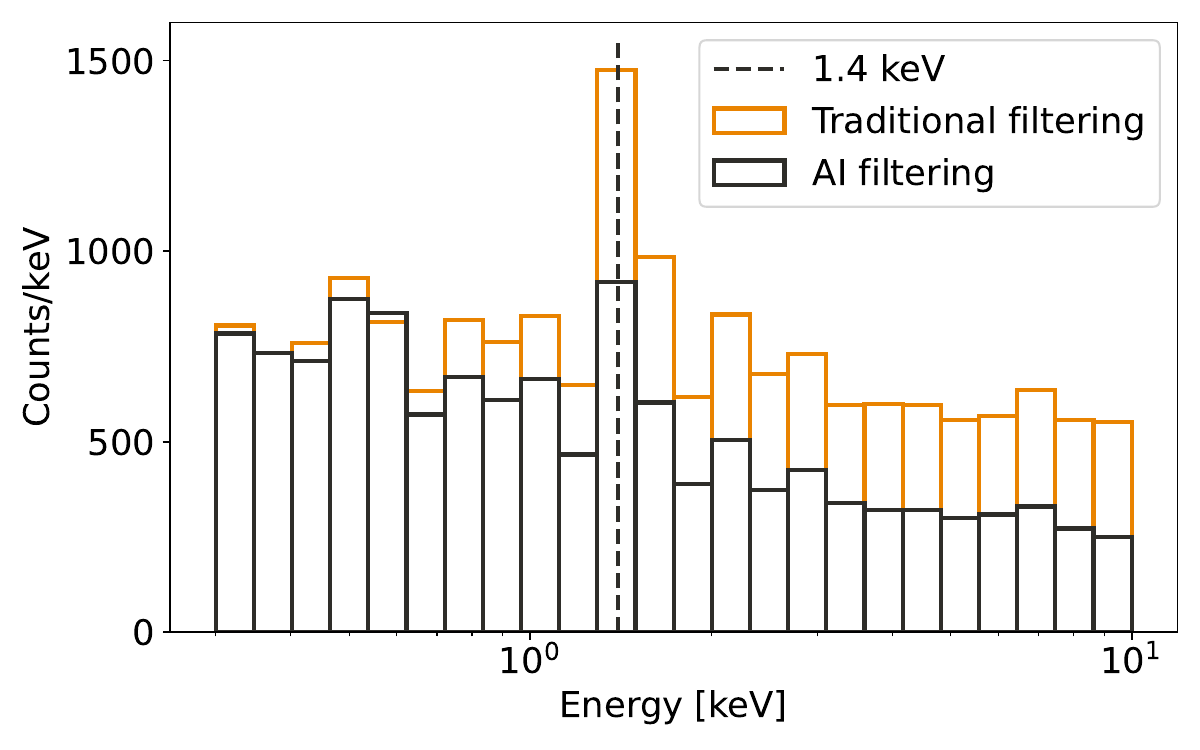}
     \caption{}
     \label{fig:unrejected_bkg}
     \end{subfigure}
    \begin{subfigure}[b]{0.49\textwidth}
    \centering
    \includegraphics[width=\textwidth]{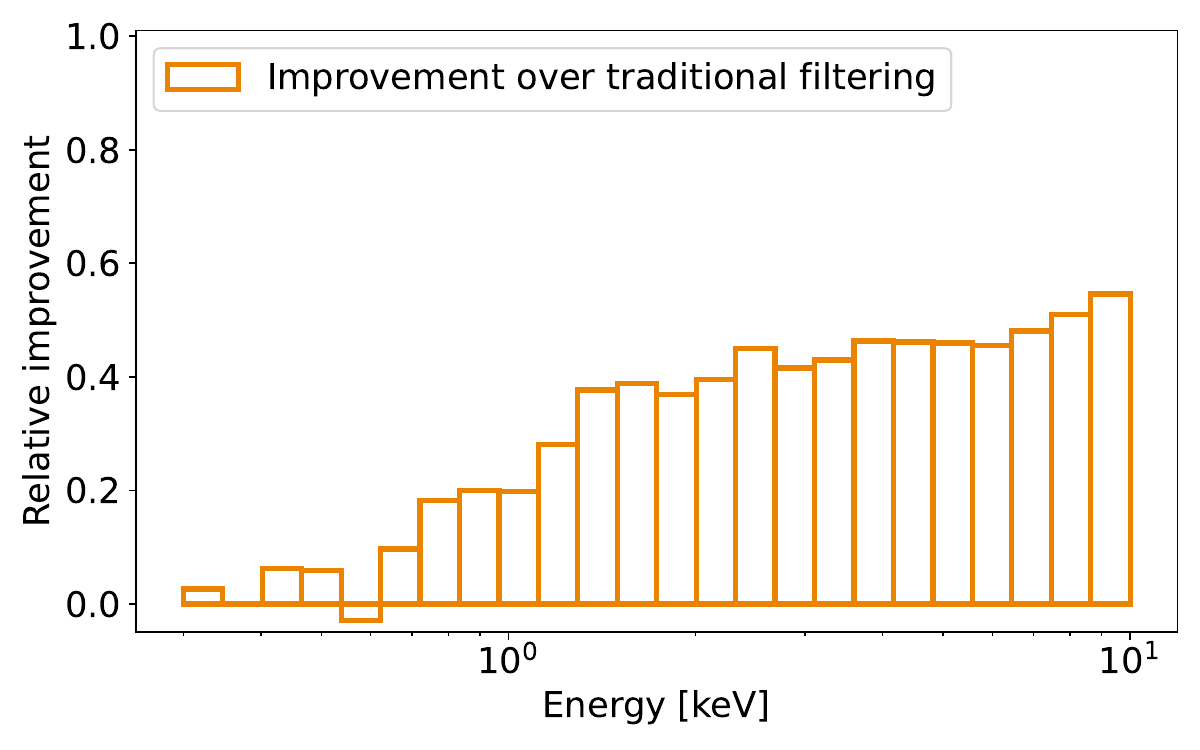}
    \caption{}
    \label{fig:unrej_bkg_rel_imp}
    \end{subfigure}
    
    \caption{Comparison of cosmic ray background rejection between traditional and AI-based filtering algorithms in 0.3-10~keV energy range. \textit{Panel a}: Spectra of unrejected background. \textit{Panel b}: Histogram of relative improvement in background rejection.}
    \label{fig:unrejected_bkg_figures}
\end{figure}

For the second stage random forest classifier we used a set of ten features obtained simultaneously with event reconstruction process. 
Beside event energy and neural network probabilities for X-ray and GCR classes, we also incorporated CAM-based features. 
These were obtained from the 5$\times$5 pixel region around the central pixel of the event. 
We used maximal value of X-ray and cosmic ray CAMs as well as median value of cosmic ray CAM. Moreover we used additional four features created out of CAM values:

\begin{equation}
\begin{aligned}
  \text{f$_{1}$} &= \text{max(CAM$_{CR}$)} - \text{min(CAM$_{X-RAY}$)} \\
  \text{f$_{2}$} &= \dfrac{\text{max(CAM$_{CR}$)}}{\text{max(CAM$_{X-RAY)}$}} - 
  \dfrac{\text{min(CAM$_{CR}$)}}{\text{min(CAM$_{X-RAY}$)}} \\
  \text{f$_{3}$} &= \dfrac{\text{max(CAM$_{CR}$)}}{\text{max(CAM$_{X-RAY)}$}} - 
  \dfrac{\text{min(CAM$_{CR}$)}}{\text{max(CAM$_{X-RAY}$)}} \\
  \text{f$_{4}$} &= \dfrac{\text{max(CAM$_{CR}$)}}{\text{min(CAM$_{X-RAY)}$}} - 
  \dfrac{\text{min(CAM$_{CR}$)}}{\text{max(CAM$_{X-RAY}$)}},
\end{aligned}
\end{equation}

\noindent
where e.g. min(CAM$_{X-RAY)}$) means a minimum value of X-ray class activation map in the 5$\times$5 pixel window centered at the most energetic pixel localized during event reconstruction.
All the CAM-based features were selected by visual inspection of the separation between classes and requirement of correlation between features being below 0.85. 
We trained random forest classifier with 1000 decision trees ensemble with the Gini impurity minimization. The number of randomly selected features a tree could use to find an optimal split in the node was set to three. 

To find an optimal probability threshold for the random forest classifier we used ROC curve (see Sec.~\ref{sec:second_stage}) shown in Fig.~\ref{fig:roc_curve}. 
The more ROC curve of the classifier deviate from the diagonal line set up by completely random classification (shown by the red line) the more effective is the classification model.
We decided to choose cosmic ray class probability threshold to be 0.77, what corresponds to 1.82\% of lost X-ray signal and 43.7\% of cosmic ray event completeness.
One can further improve classifier and stretch model's ROC curve towards upper left corner by the model hyperparameter tuning. 
This procedure as well as more detailed search for optimal second stage classifier will be discussed in the future work. 
It is worth noting that dividing mitigation pipeline into two parts makes our model a more versatile tool. 
Freedom in setting random forest rejection threshold to signal-purity or signal-completeness oriented setup allows for a tunable algorithm that can suit specific scientific needs.

\begin{figure}[t]
    \centering
    \includegraphics[width=0.49\textwidth]{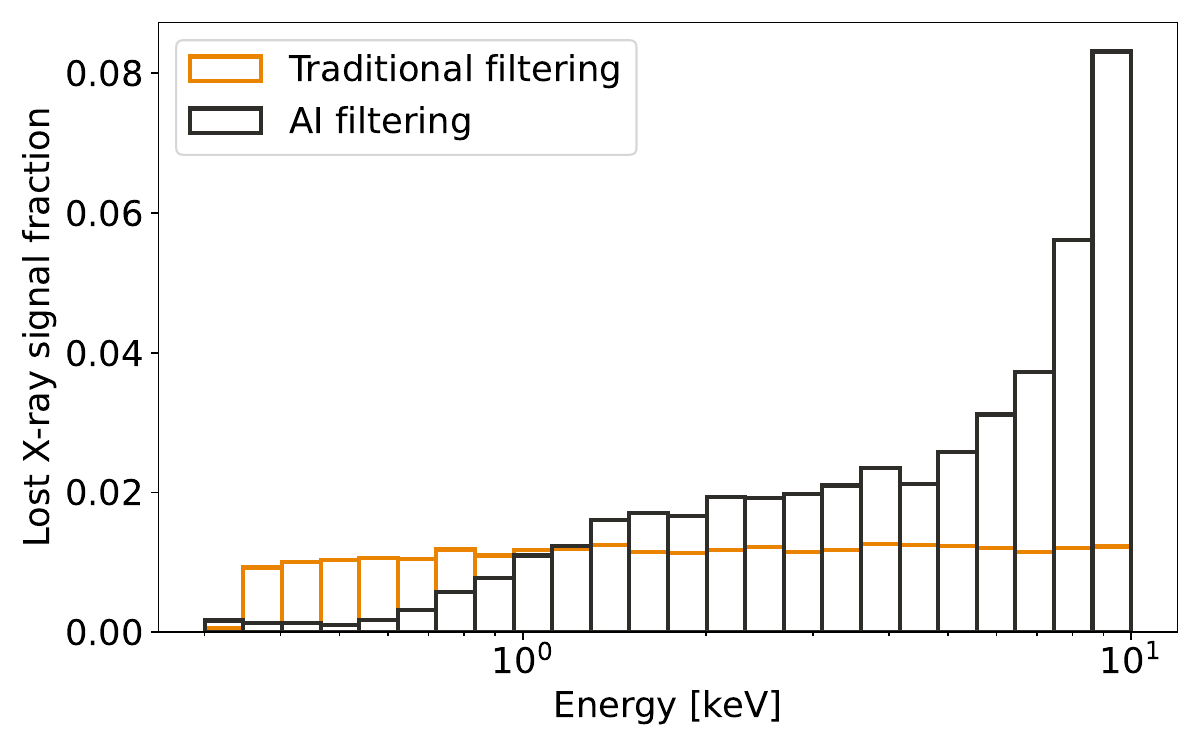}
    \caption{Comparison of lost X-ray signal level during cosmic ray background rejection between traditional and AI-based filtering algorithms in 0.3-10~keV energy range.}
    \label{fig:lost_signal}
\end{figure}

Table~\ref{tab:classification_metrics_2nd_stage} 
shows performance comparison between the second stage ML algorithm and a traditional filtering approach. 
Here we can see a significant improvement in most of the metrics. In particular we can see ~41\% of relative improvement in the level of unrejected cosmic background in 0.3-10~keV range. This improvement comes at the expense of the lost X-ray signal increase from 1.07\% to 1.82\%. 
This trade-off can be modified if needed by changing the rejection threshold of random forest classifier. 
Same improvement can be seen in the visualization of these results: Fig.~\ref{fig:unrejected_bkg_figures} shows spectrum of the unrejected cosmic ray background (Fig.~\ref{fig:unrejected_bkg}) and the relative improvement of AI filtering over traditional filtering in different energy ranges (Fig.~\ref{fig:unrej_bkg_rel_imp}).
The AI filtering shows better improvement in the higher energy range. It comes at the expense of the higher level of the lost X-ray signal as it is shown in Fig.~\ref{fig:lost_signal}, where classifier performs very aggressive cut near the 10~keV range, where cosmic ray events become more frequent. 
Other metrics focused on the performance on cosmic ray class show even more significant improvement at the level of order of magnitude  in both completeness and precision of cosmic ray class. 
When analyzing energy range of 4-8~keV, useful for iron K line science, AI filtering obtains 45.13\% of relative improvement over traditional filtering. 
This result is achieved with increase in the lost X-ray signal level to 3.11\%, while traditional filtering remains at the similar level of 1.22\%.

\section{SUMMARY}
\label{sec:summary}

In this work, we have presented the latest results from our ongoing development of machine learning methods to provide improved cosmic ray background rejection for astronomical X-ray imaging detectors.  
We showed that it is possible to obtain very effective cosmic ray localization with class activation maps (CAM) provided by a frame classification neural network. This approach allows one to use the same model for both frame classification and event localization, sidestepping the problem of the (highly) sparse data intrinsic to this problem, which challenges segmentation models.
Features extracted from CAMs contain contextual information regarding relative positions of events in the frame and their classes.
This information is used in the second stage of our pipeline to classify reconstructed events with a random forest algorithm. 
Our new method gives freedom to the user in setting the desired rejection threshold at the final stage of the pipeline, allowing a trade-off between the purity and completeness of the astrophysical X-ray signal dependent on the specific scientific needs.
For example, our approach returns a $\sim$41\% of relative improvement in cosmic ray background rejection in the 0.3-10~keV energy range compared to traditional filtering algorithms, while keeping the lost valid X-ray signal below $2\%$.

\acknowledgments 

This work has been supported by the US \textit{Athena Wide Field Imager} Instrument Consortium under NASA grant
\newline
80NSSC21M0046, and the NASA \textit{Astrophysics Research and Analysis} (APRA) program under grant number 
\newline
80NSSC22K0342. Some of the computing for this project was performed on the Stanford Sherlock cluster. We would like to thank the Stanford University and the Stanford Research Computing Center for providing computational resources and support that contributed to these research results.

\appendix
\section{SOFTWARE}
Our code was written using the Python 3 language~\cite{Python}. We used the following packages: Astropy~\cite{Astropy3, Astropy2, Astropy1}, Numpy~\cite{Numpy}, Pandas~\cite{pandas}, SciPy~\cite{SciPy}, scikit-learn~\cite{scikit-learn}, TensorFlow~\cite{tensorflow2015-whitepaper},  Matplotlib~\cite{matplotlib}.

\bibliography{main} 
\bibliographystyle{spiebib} 

\end{document}